\documentclass[conference]{IEEEtran}
\usepackage{theorem} \usepackage{cite} \usepackage{stfloats} \usepackage{epsfig} \usepackage{verbatim}
\usepackage{graphicx} \usepackage{amssymb} \usepackage{amsmath} \usepackage{color}
\usepackage[normalem]{ulem}
\usepackage{bm}
\usepackage{xcolor}
\usepackage{algorithm}
\usepackage{algorithmicx}
\usepackage{algpseudocode}
\usepackage{amsfonts}
\usepackage{cases}
\usepackage{textcomp}
\DeclareMathOperator*{\argmax}{argmax}
\ifCLASSOPTIONcompsoc
	\usepackage[caption=false,font=normalsize,labelfont=sf,textfont=sf]{subfig}
\else
	\usepackage[caption=false,font=footnotesize]{subfig}
\fi
\def\BibTeX{{\rm B\kern-.05em{\sc i\kern-.025em b}\kern-.08em
    T\kern-.1667em\lower.7ex\hbox{E}\kern-.125emX}}

\begin{document}

\title{An Efficient CSI Acquisition Method for Intelligent Reflecting Surface-assisted mmWave Networks
}

\author{\IEEEauthorblockN{Yaoshen Cui}
\IEEEauthorblockA{\textit{School of Electronic Information and Communication} \\
\textit{Huazhong University of Science and Technology}\\
Wuhan, China \\
Email: yaoshen\_cui@hust.edu.cn}\\
\and
\IEEEauthorblockN{Haifan Yin}
\IEEEauthorblockA{\textit{School of Electronic Information and Communication} \\
\textit{Huazhong University of Science and Technology}\\
Wuhan, China \\
Email: yin@hust.edu.cn}
}

\maketitle

\begin{abstract}
Millimeter-wave (mmWave) communication is one of the key enablers of the fifth-generation cellular networks (5G). However, one of the fundamental challenges of mmWave communication is the susceptibility to blockage effects. One way to alleviate this effect is the use of Intelligent Reflecting Surface (IRS). Nevertheless, due to the large number of the reflecting elements on IRS, the channel estimation turns out to be a challenging problem. This paper proposes a low-complexity channel information acquisition method for IRS-assisted mmWave communication system. Our idea consists in exploiting the sparsity of the mmWave channel and the topology of IRS itself. Compared to the state-of-the-art methods, our proposed method requires much less time-frequency resource in the acquisition of channel information. Large performance gains are confirmed in simulation, which prove the effectiveness of our proposed method.
\end{abstract}
\begin{IEEEkeywords}
IRS, LIS, metasurface, mmWave, blockage, coverage, positioning
\end{IEEEkeywords}

\section{Introduction}
In the past year, 5G commercialization is in full swing, with the fact that the number of devices and the volume of data continue to grow exponentially which challenges the current communication networks. The first commercial use of 5G technology is in general believed to be the enhanced mobile broadband (eMBB), which requires much higher throughput than the forth generation (4G) networks. In fact, the ever-increasing data rate is always a major requirement in cellular networks. In order to meet this requirement, a variety of technologies have been proposed and thoroughly investigated in the last decade, including ultra-dense networking (UDN), massive multiple-input multiple-output (massive MIMO) and millimeter wave (mmWave) communications\cite{F.Boccardi:2014Five}.

Among them, the application of mmWave will bring a critical improvement to 5G, by offering an enormous amount of spectrum resource that is an order of magnitude more than 4G.
Although the absorption due to air and rain  does not create significant additional path loss for cell sizes on the order of 200 meters for mmWave, the frequencies of that is much more sensitive to blockages compared with microwave\cite{TS:2013MmwaveWork}, which undermines the potential of mmWave. Thus, the coverage problem of mmWave have to be addressed in order to realize the practical value of the large bandwidth.

From the above problem of blockage, we realized the lack of control of the wireless propagating environment. Fortunately, a concept of IRS\cite{marco:2019metasuface}\cite{wuqq:2019IRS} (also known as metasurface\cite{Tang:2019meta}, large intelligent surface (LIS)\cite{Hu:2018LIS}\cite{Huang:2018LIS}) is presented timely, which has the capability of intelligently reflecting radio waves. Moreover, IRS is a cost-effective solution due to the low hardware cost  and energy consumption. In principle, the full potential of IRS can be realized when the channel state information (CSI) between the IRS and the wireless transceiver is known. However, since the number of reflecting units in IRS is very large, and they are all passive and unable to sense the signal, the CSI acquisition is a challenging problem. Several attempts have been made to address this problem. \cite{He:2019EstLIS} proposed a three-stage cascaded channel estimation framework. \cite{zheng:2019intelligent} proposed a transmission protocol and successive reflection optimization method. However, the common problem of the known methods is the resource consumption in channel training.

In this paper, we propose an efficient channel acquisition method for the IRS-assisted mmWave communication system. The key of our method relies on 1) the geometry of the metasurface, and 2) the structural information of the wideband wireless channels. More specifically, we propose a concept named reflecting unit set (RUS), which consists of a subset of neighboring reflecting units. The positions of the RUSs are properly selected from the metasurface. Then the line-of-sight (LOS) distance between a certain RUS and a user terminal (UE) can be estimated by choosing an optimal codeword of phase shifts that maximizes the received power on UE side. With several RUS-UE distances, we are able to reconstruct the channel based on the structure of the multipath response. Our method has much lower computational complexity as well as the time-frequency resource consumptions for channel acquisition compared with the state-of-the-art methods, as our method is able to learn the channel within around 20 OFDM symbols (or 0.167 ms with 120 kHz of subcarrier spacing for mmWave \cite{3gpp:38.211}) even in the presence of several thousands of reflecting units. To the best of our knowledge, this is the first low-cost CSI acquisition method for IRS-assisted network based on the RUS concept.

\emph{Notations:} We use bold-face to denote vectors and matrices. $\mathbb{C}^{x\times y}$ denotes the space of $x\times y$ complex-valued matrices. $\odot$ stands for the Hadamard product. For a vector $\bm{x}$, $\|\bm{x}\|$ denotes its the Euclidean norm. $\bm{X}^T$, $\bm{X}^*$, $\bm{X}^H$ denote the transpose, conjugate and  conjugate transpose of a matrix $\bm{X}$, respectively.

\section{Signal and channel model}\label{sec:model}
\begin{figure}[!t]
\centering
\includegraphics[width=3.5in]{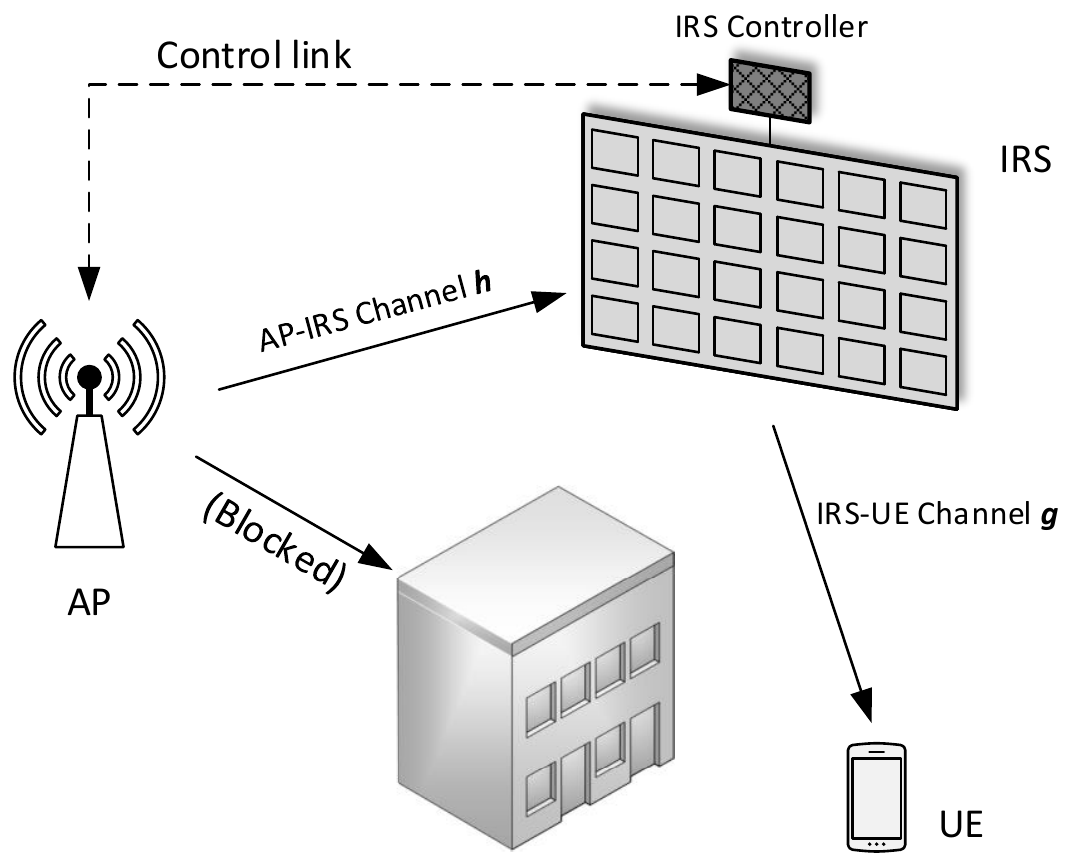}
\caption{IRS-assisted Communication System}
\label{fig:IRS-aided Communication System}
\end{figure}

We consider a mmWave communication system with blockages in the propagation environment. Due to the high frequency of mmWave, the transmitted signal from an Access Point (AP) to a User Equipment (UE) is  often blocked by obstacles, since the diffraction effects are negligible and the number of scattering paths is much smaller than in lower frequency bands (e.g., below 3 GHz). In this case, an IRS is deployed to assist the communications. An illusion of the system is shown in Fig. \ref{fig:IRS-aided Communication System}.

The number of reflecting units at the IRS is denoted by $N$. Let $\bm{\phi}=[\phi_{1},\ldots,\phi_{\emph{N}}]$ and $\bm{\beta}=[\beta_{1},\ldots,\beta_{\emph{N}}]$, where $\phi_{n}\in[0,2\pi)$ and $\beta_{n}\in[0,1], (n = 1,\ldots,N)$  denote the phase shift and the amplitude reflection coefficient of the $n$-th reflection unit of the IRS, respectively. The reflection-coefficients vector of the IRS is expressed as
\begin{equation}\label{eq:reflection-coefficients matrix of the IRS}
\begin{split}
\bm{\theta}&={[\theta_{1},\cdots,\theta_{\emph{N}}]}^T\\
           &={[\beta_{1}e^{j\phi_{1}},\ldots,\beta_{\emph{N}}e^{j\phi_{\emph{N}}}]}^T,
\end{split}
\end{equation}
where $\theta_{n}=\beta_{n}e^{j\phi_{n}}, (n=1,\ldots,N)$ denotes the reflection coefficient of the $n$-th reflection unit  of the IRS. Note that in practice, each unit on the IRS are designed to maximize signal reflection. Thus, we set $\beta_n = 1, n = 1,\ldots,N$ in the sequel of this paper.

For ease of exposition, we consider a single transmit and receive antenna, yet our method can be readily generalized to multiple-antenna AP/UE setting. The transmitted signal at the AP is denoted by $x$. The AP-IRS channel and IRS-UE channel are denoted by $\bm{h}\in\mathbb{C}^{1\times N}$ and $\bm{g}\in\mathbb{C}^{1\times N}$, respectively. Thus, the signal received at the user is expressed as
\begin{equation}\label{eq:y signal recevied a}
y=(\bm{g}\odot\bm{h})\bm{\theta}x+z,
\end{equation}
where $z \sim \mathcal{CN}(0,\sigma^2)$ denotes the additive white Guassian noise (AWGN) at the UE side.

Then, the LOS channel between the AP and the $n$-th reflecting unit, denoted by $h_{n}$, can be expressed as \cite{3gpp:38.901}:
\begin{equation}
h_{n}=\rho_{\emph{ap,n}}e^{\frac{-j2\pi fd_{\emph{ap,n}}}{c}},\label{eq:AP-n channel}
\end{equation}
where $\rho_{\emph{ap,n}}$, $f$, $c$, $d_{\emph{ap,n}}$ denote the path loss between the AP and the $n$-th reflecting unit, the frequency, the speed of light, and the distance between AP and the $n$-th reflecting unit.

Note that the path loss $\rho$ is based on the distance between the $n$-th reflecting unit and the AP/UE, which can be expressed as\cite{yin:13a}
\begin{equation}\label{eq:path loss}
\rho=\frac{\alpha}{d^{\gamma}},
\end{equation}
where $\alpha$ is a constant dependent on the precribed signal-to-noise-ratio (SNR), $d$ is the geographical distance, $\gamma$ is the path-loss exponent. Thus, the path loss between the AP and the $n$-th reflecting unit can be expressed as
\begin{equation}\label{eq:AP-n path loss}
\rho_{\emph{ap,n}}=\alpha/d_{\emph{ap,n}}^{\gamma}.
\end{equation}

Based on (\ref{eq:AP-n channel}) and (\ref{eq:AP-n path loss}), the AP-IRS channel can be expressed as
\begin{equation}\label{eq:AP-IRS channel}
\begin{split}
\bm{h}&=[h_{1},\cdots,h_{\emph{N}}]\\
	&=\begin{bmatrix}
	\frac{\alpha}{d_{\emph{ap,}1}^{\gamma}}e^{\frac{-j2\pi f d_{\emph{ap,}1}}{c}}, \cdots,
	\frac{\alpha}{d_{\emph{ap,N}}^{\gamma}}e^{\frac{-j2\pi f d_{\emph{ap,N}}}{c}}
	\end{bmatrix}.
\end{split}
\end{equation}

Similarly, the channel between the $n$-th reflecting unit and the UE can be written as:
\begin{align}
g_{n}&=\rho_{\emph{n,ue}}e^{\frac{-j2\pi fd_{\emph{n,ue}}}{c}},\label{eq:n-UE channel}\\
\rho_{\emph{n,ue}}&=\alpha/d_{\emph{n,ue}}^{\gamma},\label{eq:n-UE path loss}
\end{align}
where $\rho_{\emph{n,ue}}$ and $d_{\emph{n,ue}}$ denote respectively the path loss and distance between the $n$-th reflecting unit and the UE. Thus, the  IRS-UE channel, denoted by $\bm{g}$, can be expressed as
\begin{equation}\label{eq:IRS-UE channel}
\begin{split}
\bm{g}&=[g_{1},\cdots,g_{\emph{N}}]\\
	&=\begin{bmatrix}
	\frac{\alpha}{d_{1\emph{,ue}}^{\gamma}}e^{\frac{-j2\pi fd_{1\emph{,ue}}}{c}}\cdots,
	\frac{\alpha}{d_{\emph{N,ue}}^{\gamma}}e^{\frac{-j2\pi fd_{\emph{N,ue}}}{c}}
	\end{bmatrix}.
\end{split}
\end{equation}

In order to combat the blockage effects, we seek to maximize the received power at UE side by adjusting the reflecting coefficient vector $\bm{\theta}$ based on the knowledge of CSI, which can be expressed as
\begin{align}
&\bm{\theta}=\argmax\limits_{\bm{\theta}}\{|(\bm{g}\odot\bm{h})\bm{\theta}|^2\}\label{eq:opt Theta a}\\
&\rm{s.t.} \quad |\theta_{n}| \le 1,n=1,\cdots,N.
\end{align}

Ideally, when the CSI is known, the optimal reflecting coefficient of the $n$-th reflecting unit can be expressed as
\begin{equation}\label{eq:normalize theta}
\theta_{n}=\frac{g_n^*h_n^*}{|g_nh_n|},n=1,\cdots,N.
\end{equation}

Nevertheless, the channel information has to be estimated beforehand. Due to the large number of reflecting elements which do not have any signal processing capability, the CSI acquisition is a challenging problem. Moreover, the time-varying location of the UE makes it imperative to learn the channel within a very short time. In next section, we will propose an efficient CSI acquisition method for the mmWave channel between IRS and the transceivers.

\section{Proposed CSI acquisition method}\label{sec:proposed}
The sparse channel of mmWave is composed of one or several paths \cite{Robert:2016mmwave}. The strongest path is normally the LOS path, if it exists. Due to the large number of co-located reflecting units, the channel between the IRS and the AP/UE has a unique structure that can be utilized during channel acquisition. In this section, we try to exploit the sparsity and the structural information of the mmWave channel in order to obtain the LOS path response between the IRS and the transceivers. We may then design the phase shift vector $\bm{\theta}$ according to the LOS channel between the IRS and AP/UE in order to enhance the coverage of mmWave by circumventing the blockages between AP and UE.

\begin{figure}[!t]
\centering
\includegraphics[width=3.5in]{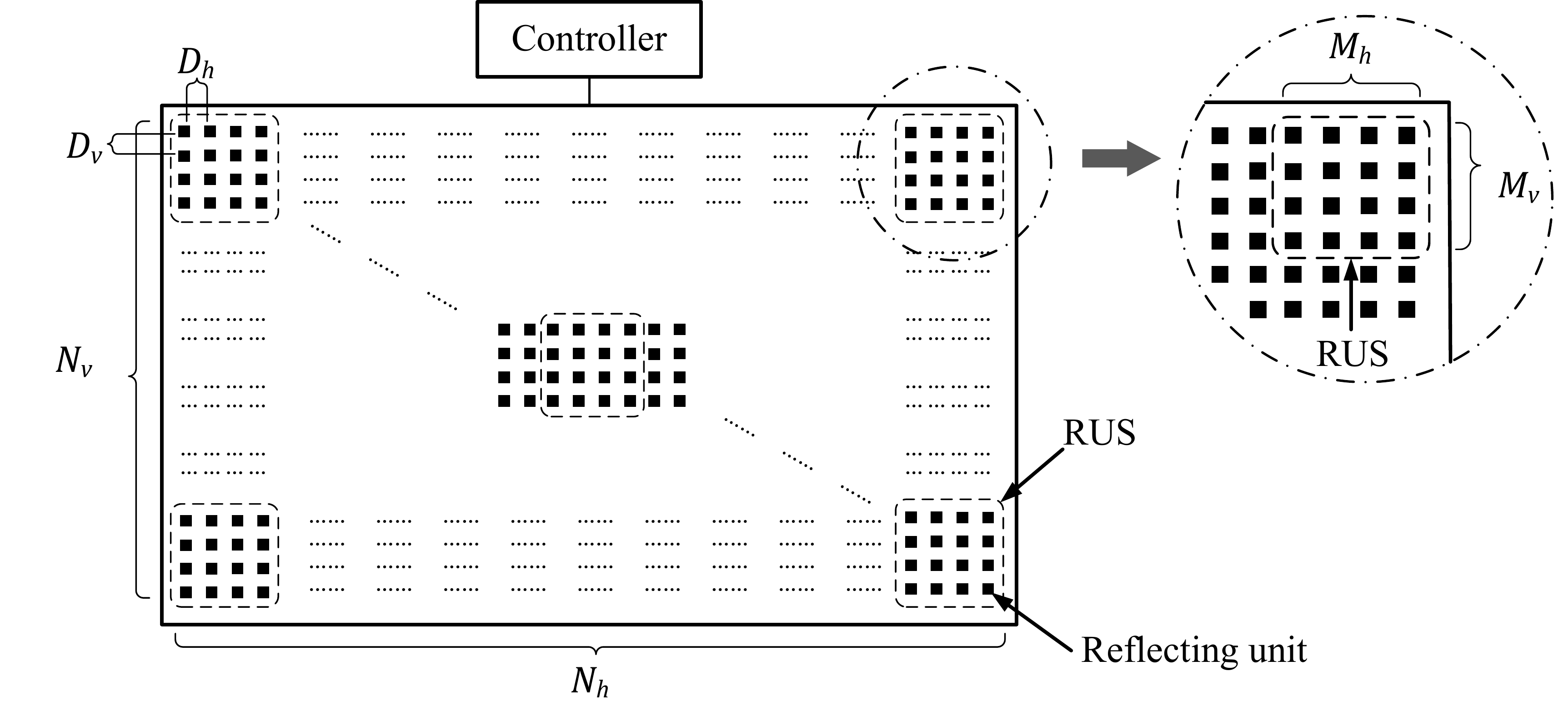}
\caption{RUS example in IRS.}
\label{fig:IRS example}
\end{figure}

The key idea of our CSI acquisition method is to locate the position of the UE by channel response of the reflected radio wave. Due to the large size of IRS, the far-field propagating model does not hold anymore, especially in a indoor setting. The distances between different reflecting units and the transceiver diverse a lot. As a result, we propose to calculate the location of UE by estimating the AP-reflecting unit-UE distances and then use triangulation algorithm \cite{wang:2013triang} to obtain the 3D position of UE. In order to achieve this, we propose a concept named reflecting unit set (RUS) which is a set of neighboring reflecting units as shown in Fig. \ref{fig:IRS example}. The purpose of RUS is to amplify the reflected signal in order for UE to detect the received signal, providing that a proper codeword of phase shifts is chosen and applied to current RUS. More details are given below.

\subsection{Reflecting Units Set}\label{sec:RUS}
By definition, RUS is a set of reflecting units in a certain part of the IRS. For ease of exposition, we make the assumption that all RUSs have the same shape and size. An RUS can be a rectangular structure with $M_v$ rows and $M_h$ columns. The total number of units of the RUS is $M = M_h M_v$. The total number of the RUSs is $N_M$. The 3D cartesian coordinate of the center of the $m$-th RUS is denoted as $\bm{P}_{m}\in\mathbb{C}^{1\times 3},m=1,\ldots,N_M $.

Then we introduce some key characteristics of RUS:
\begin{itemize}[\IEEEsetlabelwidth{$\gamma$}]
\item \emph{Active State:}
	An RUS is \emph{Active} means the reflecting elements in current RUS is in a working state, while the remaining elements on this IRS either stop reflecting (shut down) or reflect the incident wave to a direction where UE is very unlikely to exist.

\item \emph{Codebook and Codeword:}
    The codebook of an RUS is a set of codewords, which are used for reflecting the radio wave to different directions. Each codeword is a column vector of size $M \times 1$, with the $m$-th value being the reflecting coefficient for the $m$-th elements in the RUS. Since the RUS is selected as a uniform planar array (UPA), we may adopt the discrete Fourier transform (DFT) based codebook as defined in \cite{3gpp:38.214}:
\begin{small}
\begin{align}
u_{p}&= \begin{bmatrix}1 \quad e^{j\frac{2\pi p}{O_1 M_v}} \quad \cdots \quad e^{j\frac{2\pi p(M_v-1)}{O_1M_v}} \end{bmatrix},\\
v_{l,p}&=\begin{bmatrix}u_p \quad e^{j\frac{2\pi l}{O_2M_h}}u_p \quad \cdots \quad e^{j\frac{2\pi l(M_h-1)}{O_2M_h}}u_p \end{bmatrix}^{T},
\end{align}
\end{small}
The codewords are $v_{l,p}$ with $p = 1, \cdots, M_v$ and $l = 1, \cdots, M_h$ being the indices of vertical and horizontal beam directions respectively. $O_1$ and $O_2$ are the vertical and horizontal oversampling factors that may take positive integer values. We can choose from the codebook a codeword that maximizes the received signal at UE. It may take several trials to find a good codeword.

\end{itemize}

Note that the shape of the RUS is not limited to a UPA, and the size of the RUS can be customized. Smaller RUS may have better accuracy when the reflected signal is strong enough for UE to detect. However, when the strength of the desired reflective radio wave is not guaranteed, the size of the RUS should be larger.

We then show the RUS-based UE position estimation method below.
In order to estimate the UE's position with triangulation algorithm, two key parameters are needed: the positions of the observation points and the distances between UE and the corresponding observation points. In this paper, we select the RUSs as the observation points and set the centers of RUSs as their positions. The RUS-UE distance can be obtained using classical wideband delay estimation method, for example the Bartlett's method \cite{EH:2012bartlett}, as shown below.

\subsection{Bartlett-based Estimation of Delays}\label{sec:multi-subbands}
The wideband channel sounding signal in 5G can be utilized to estimate the delay of the AP-RUS-UE channel. The whole bandwidth is equally divided into $K$ subbands, where the $k$-th subband has a center frequency of $f_k$. When the $m$-th RUS is activated and a proper codebword is chosen, we may have an estimate of the channel of all $K$ subbands, denoted by $\mathbf{\hbar}_m$:
\begin{equation}\label{Eq:wbche}
\bm{\hbar}_m = [\hbar_m(f_1), \hbar_m(f_2), \ldots, \hbar_m(f_K)],
\end{equation}
where $\hbar_m(f_k), k = 1, \cdots, K$ is the channel at the $k$-th subband. The delay of the AP-RUS-UE channel when the $m$-th RUS is activated can be calculated as
\begin{equation}\label{eq:delay of AP-RUS-UE}
t=\argmax_{t}\{|\bm{\hbar}_m\bm{b}(t)|^2\},
\end{equation}
where $\bm{b}(t)\in\mathbb{C}^{K\times 1}$ is defined as
\begin{equation}\label{eq:b(t)}
\bm{b}(t)=[e^{j2\pi f_1t},\cdots,e^{j2\pi f_Kt}]^T.
\end{equation}
Note that some other delay estimation algorithms are also available, e.g., the Multiple Signal Classification (MUSIC) \cite{R.schmidt:1986music}\cite{Nikolic:2013music}, Estimation of Signal Parameters via Rational Invariance Techniques (ESPRIT) \cite{roy:1989esprit}\cite{Duofang:2008esprit}, etc.

\subsection{Triangulation Positioning Algorithm}\label{sec:tri-alg}
\begin{figure}[!t]
\centering
\includegraphics[width=3in]{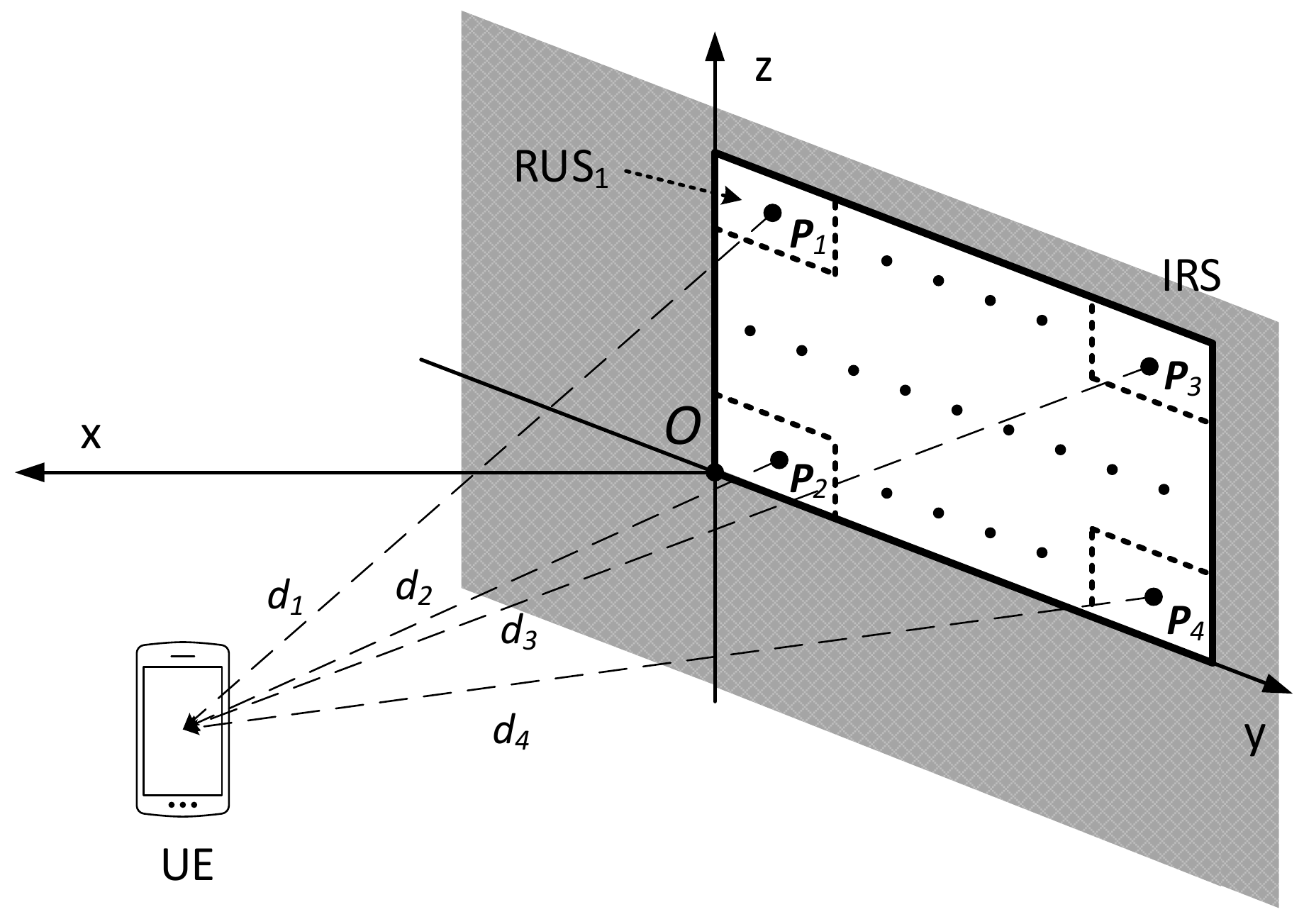}
\caption{IRS-based UE position triangulation}
\label{fig:tri show}
\end{figure}

As shown in Fig. \ref{fig:tri show}, we establish a 3D coordinate system in which the reflecting unit in the lower left corner of the IRS is the origin. The IRS is in the $y$-$z$ plane. The AP and UE are in the same side of the IRS, so that the $x$-coordinates of both AP and UE are positive.

The positions of the RUSs in the coordinate system are assumed known. The location of the $m$-th RUS is denoted by $\bm{P}_m = (x_m, y_m, z_m), m = 1, \cdots, N_M $. The distances between the $m$-th RUS ($m = 1, \cdots, N_M$) and the UE, denoted by $\hat{d}_m$ has been computed by the delay estimation in Sec. \ref{sec:multi-subbands}. The estimated position of UE, denoted by $\bm{\hat{P}} = (\hat{x}, \hat{y}, \hat{z})$ can be found by solving the following equations under the constraint that $\hat{x} \geq 0$:
\begin{equation}\label{eq:triangulation dm}
\begin{split}
{\hat{d}_m}&=\|\bm{P}_m-\hat{\bm{P}}\|\\
     &=\sqrt{(x_m-\hat{x})^2+(y_m-\hat{y})^2+(z_m- \hat{z})^2},
\end{split}
\end{equation}
where $m = 1, \cdots, N_M$.

In principle, we may obtain an estimate of UE position using three RUSs and the corresponding distances. Nevertheless, we can increase the accuracy and robustness of the positioning algorithm by activating more than three RUSs.

\subsection{Proposed Strategy of Reflecting Coefficient Calculation}
\begin{figure}[!t]
\centering
\includegraphics[width=3.5in]{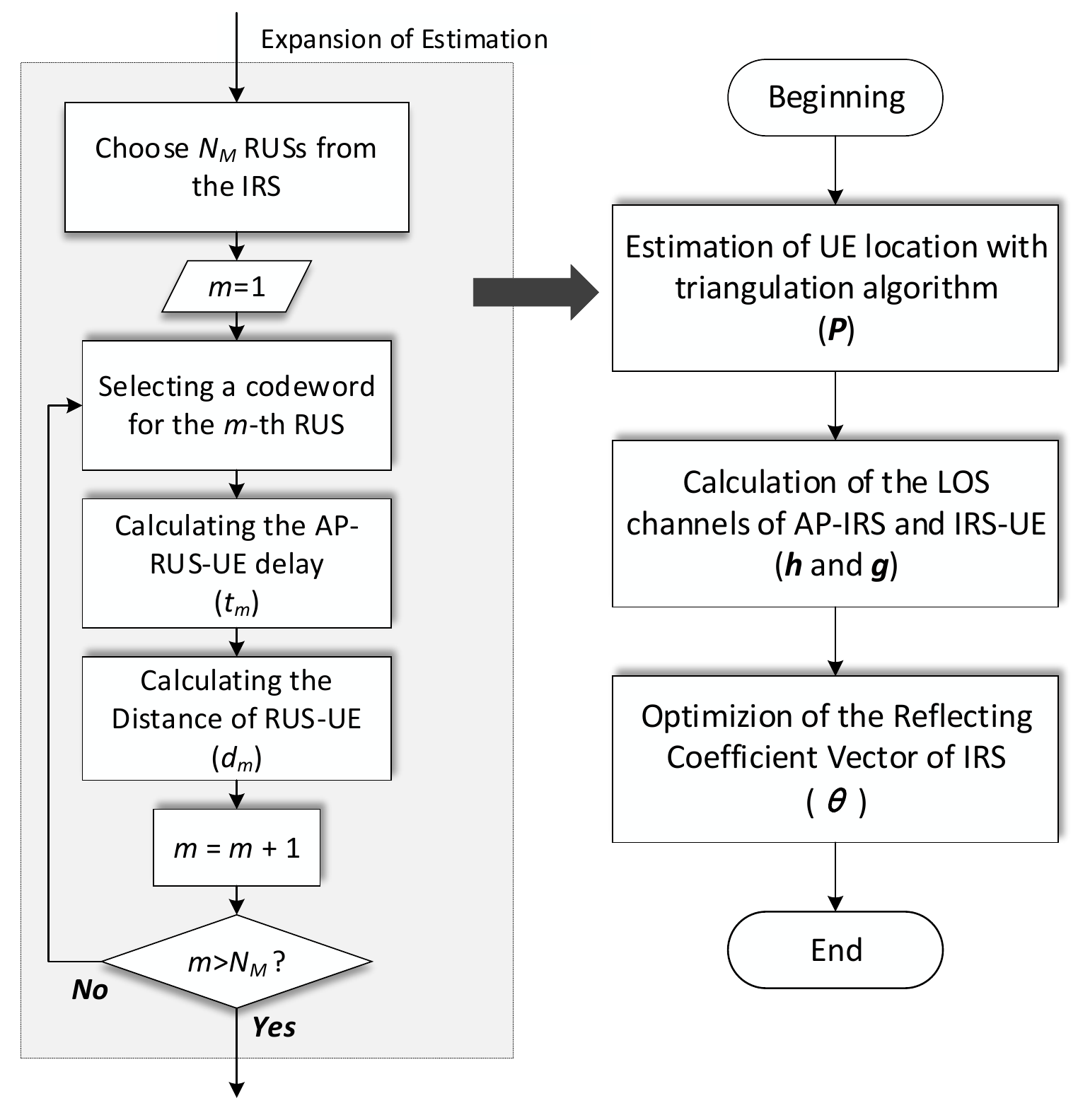}
\caption{Flow chart of the reflecting coefficient computation of IRS based on the estimation of UE position.}
\label{fig:Flow chart}
\end{figure}

With the above tools and methods, we now show our proposed method of CSI acquisition and and reflecting coefficient calculation approach. The flow chart of our method is shown in Fig. \ref{fig:Flow chart}. The core steps are listed below. The locations of AP, IRS, and RUSs are assumed known. The codebook of RUS is denoted by $\mathcal{W}$.

\begin{enumerate}[\IEEEsetlabelwidth{4}]
\item \textbf{Calculate the distance between each RUS and the UE}. Activate all RUSs sequentially and for the $m$-th RUS $(m = 1, \cdots, N_M)$, we choose a codeword $\bm{w} (m)$ from the codebook $\bm{W}$ that maximizes the received power at UE side. This operation requires the feedback from UE to indicate the power of received signal. Note that a complete search may not be necessary, as long as the reflected signal from the activated RUS can be identified by UE from noise and interference. Since the RUS is much smaller than IRS, it is also very likely that one codeword can be shared by all RUSs so that the searching might be done only once. This will greatly increase the efficiency of the algorithm. Once $\bm{w} (m)$ is determined, UE estimates its channel under $\bm{w} (m)$ using the training signal sent by AP and then calculate the AP-RUS-UE distance as shown in Sec. \ref{sec:multi-subbands}. Since the distance between AP and a certain RUS is known, we may readily obtain the distance between an RUS and UE.

\item \textbf{Calculate the estimated position of UE}. Once the RUS-UE distances are obtained, we apply the triangulation algorithm as shown in Sec. \ref{sec:tri-alg}. We denote the estimated coordinate of UE as $\widehat{\bm{P}}$.

\item \textbf{Calculate the channels from AP to IRS and from IRS to UE}. According to the channel model of Eq. (\ref{eq:IRS-UE channel}), we may reconstruct the channel between IRS and UE with the estimated UE position $\widehat{\bm{P}}$. Denote this channel estimate as $\hat{\bm{g}} = [\hat{g}_{1}, \hat{g}_{2}, \cdots, \hat{g}_{\emph{N}}]$. Since the coordinates of AP and IRS are known, we may readily obtain the AP-IRS channel $\bm{h} =[h_{1}, h_{2}, \cdots,h_{\emph{N}}]$ between AP and IRS using the model of Eq. (\ref{eq:AP-IRS channel}). Note that, in fact only the phases of $\hat{\bm{g}}$ and $\bm{h}$ are needed.

\item \textbf{Calculate the desired reflecting coefficients of IRS}. Finally, we compute the desired phase shifts on the IRS based on the channel information $\hat{\mathbf{g}}$ and $\bm{h}$. The solution is expressed as:
\begin{align}\label{eq:Theta-opt}
&\bm{\theta}_\text{opt} = \argmax\limits_{\bm{\theta}}\{|(\hat{\bm{g}}\odot\bm{h})\bm{\theta}|^2\}\\
&\rm{s.t.} \quad |\theta_{n}| \le 1,n=1,\cdots, N,
\end{align}
which yields
\begin{equation}\label{eq:Theta-opt2}
\theta_\text{opt, n}=\frac{\hat{g}_n^*h_n^*}{|\hat{g}_n h_n|},n=1, \cdots, N,
\end{equation}
where $\theta_\text{opt, n}$ is the desired coefficient of the $n$-th reflecting unit.
\end{enumerate}

The above proposed method can be completed in a very short time. Consider for example an IRS with four RUSs, each of the RUS has four rows and four columns of reflecting elements. A DFT codebook containing 16 codewords is used. Assuming the same codeword is shared by all RUSs, the searching of the codeword takes 16 OFDM symbols. Three more symbols are needed to complete activating all RUSs. Thus the total number of OFDM symbols needed can be limited within 20.

\section{Numerical Results}\label{sec:simulation}
\begin{figure}[!t]
\centering
\includegraphics[width=3.5in]{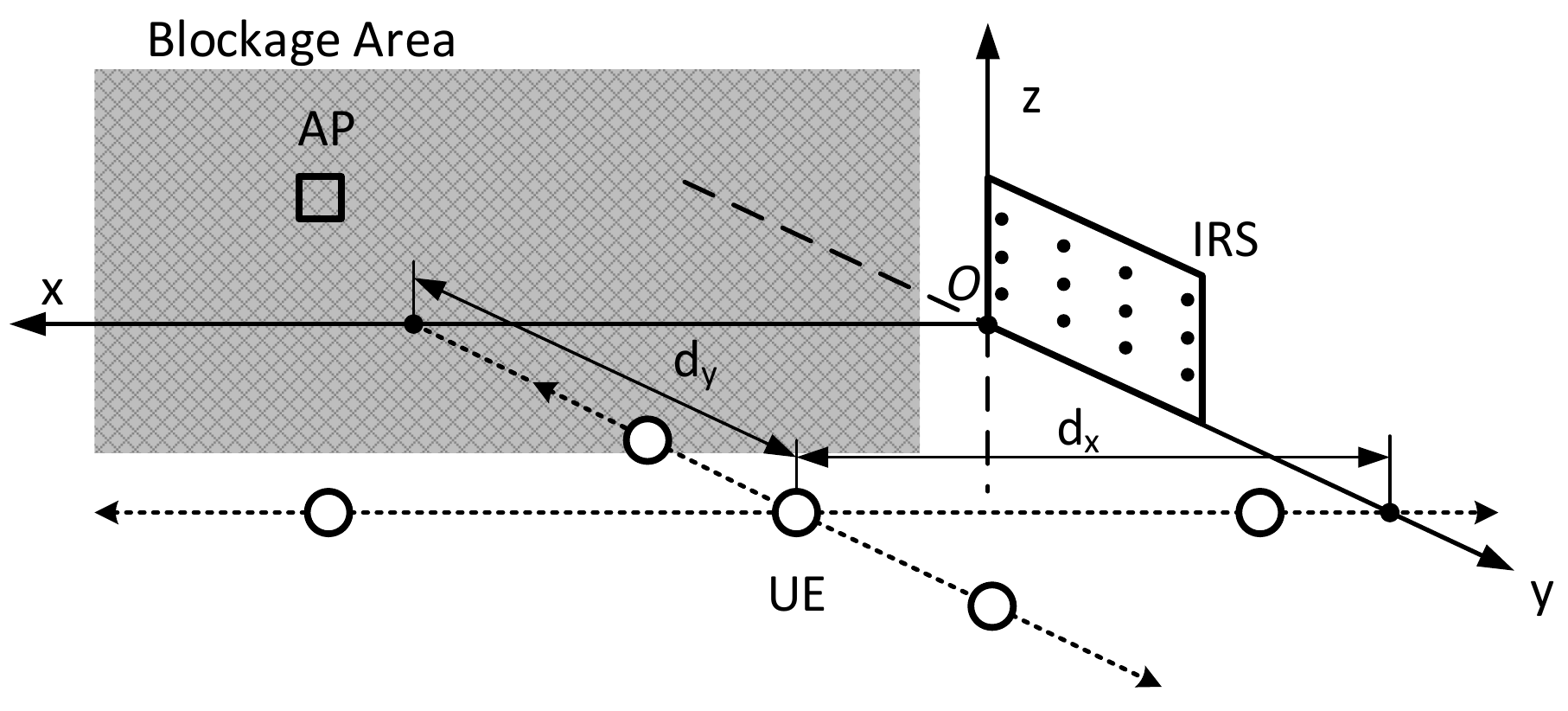}
\caption{Simulation setup (top view).}
\label{fig:simulation setup}
\end{figure}

In this section, we verify the feasibility of the proposed design through multiple simulations and analyze its performance and robustness. As shown in Fig. \ref{fig:simulation setup}, we consider a 3D coordinate system where the IRS is composed of a UPA of reflecting units. Without loss of generality, the IRS is located in $y$-$z$ plane and the reflection unit in IRS lower left corner is at the origin of the coordinate system, the unit length of which is 1 meter. The reference (center) antenna at the AP is located at $(x_{_\emph{AP}},y_{_\emph{AP}},z_{_\emph{AP}})=(5,-5,0)$.
The channel estimate in Eq. (\ref{Eq:wbche}) is corrupted by noise, which is modeled as:
\begin{equation}\label{eq:sigma-e}
\hat{\bm{\hbar}}_m=\sqrt{1-\sigma_e^2}\bm{\hbar}_m+\sigma_e\bm{z},
\end{equation}
where $\bm{z} \in\mathbb{C}^{1\times K}$ is the independent and identically distributed (i.i.d.) additive Gaussian noise with element-wise variance being 1. $\bm{\hbar}_m$ is the true channel of all $K$ subbands for the $m$-th RUS. $\hat{\bm{\hbar}}_m$ is the corresponding noisy channel estimate. $\sigma_e \in [0, 1]$ is the quality of the channel estimate.


We assume the direct channel between AP and UE is negligible due to the blockage effects.
The IRS contains $N=8192$ reflecting elements with $N_v = 64$ rows and $N_h = 128$ columns. The horizontal and vertical spacings between neighboring reflecting units are both half wavelength of the center frequency. More parameters can be found in Table \ref{tab:paras}.
We set the order of reflecting units on IRS as follows: starting from the lower left corner of UPA, that is, the origin of the coordinate, the number increases along the first column, then the second, etc, until the $N_h$-th column.

\begin{figure}[!t]
\centering
\includegraphics[width=3.2in]{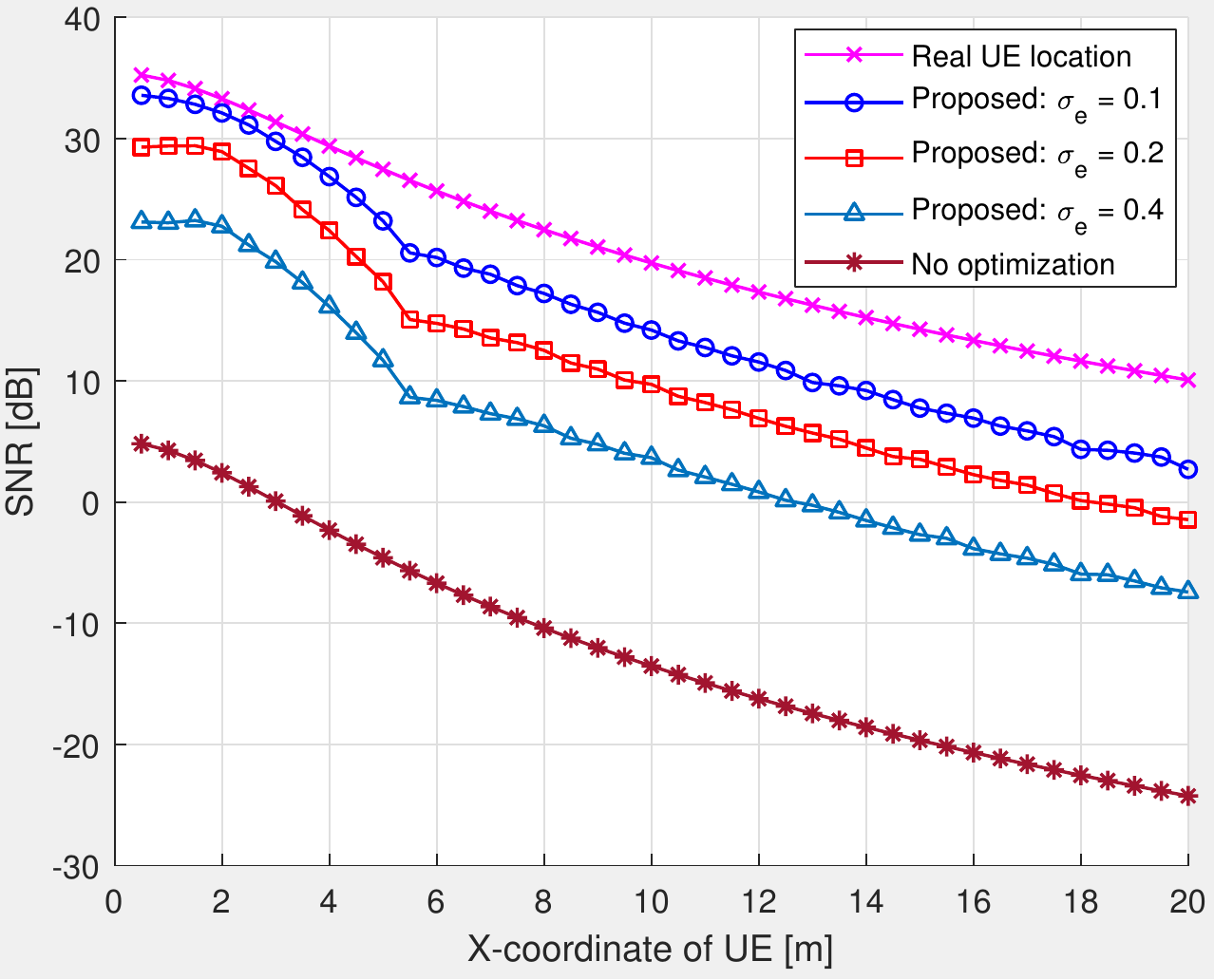}
\caption{Recived SNR vs. X-coordinate of UE}
\label{fig:sim x snr}
\end{figure}

\begin{figure}[!t]
\centering
\includegraphics[width=3.2in]{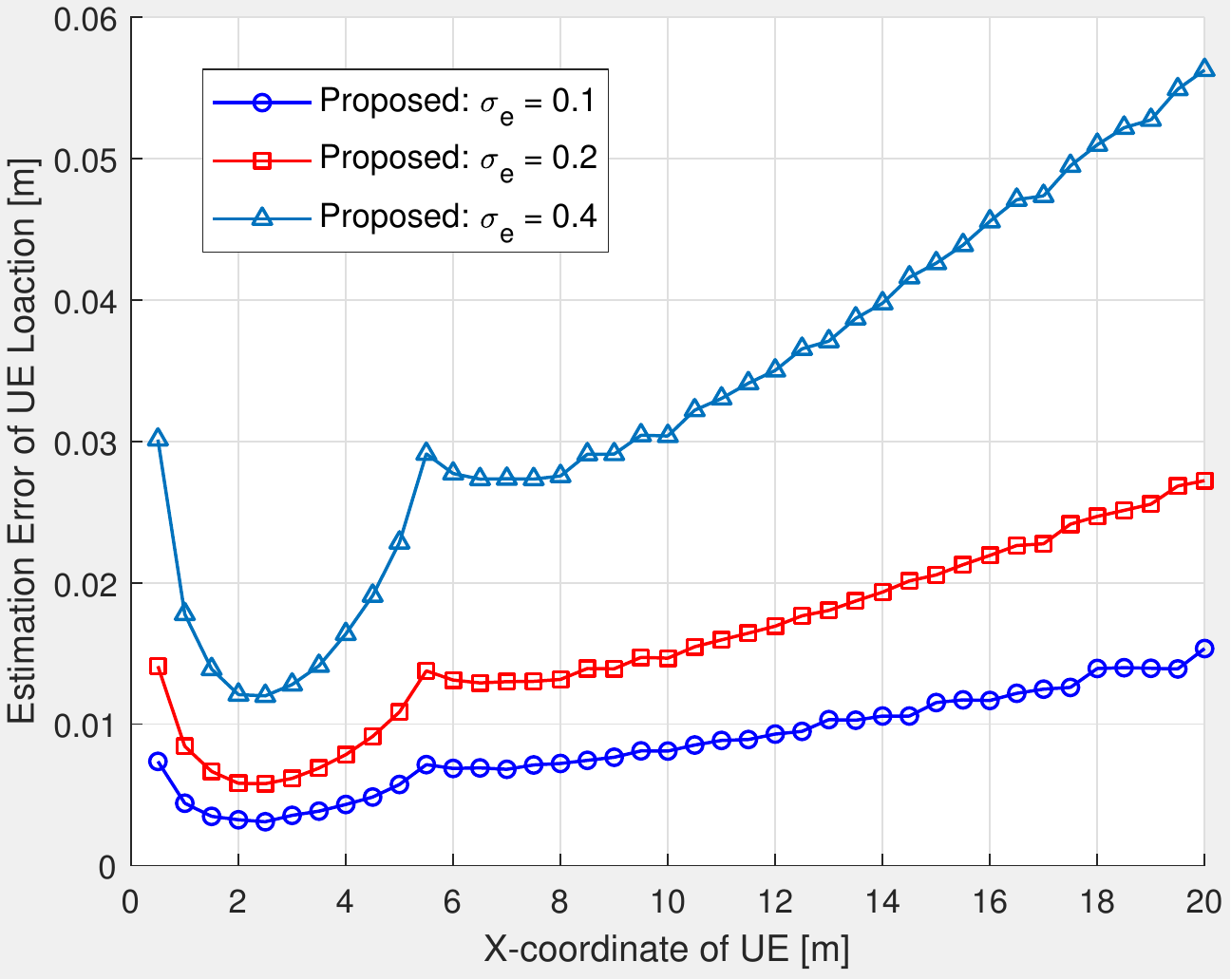}
\caption{Location Estimation Error vs. X-coordinate of UE}
\label{fig:sim x err}
\end{figure}

\begin{figure}[!t]
\centering
\includegraphics[width=3.2in]{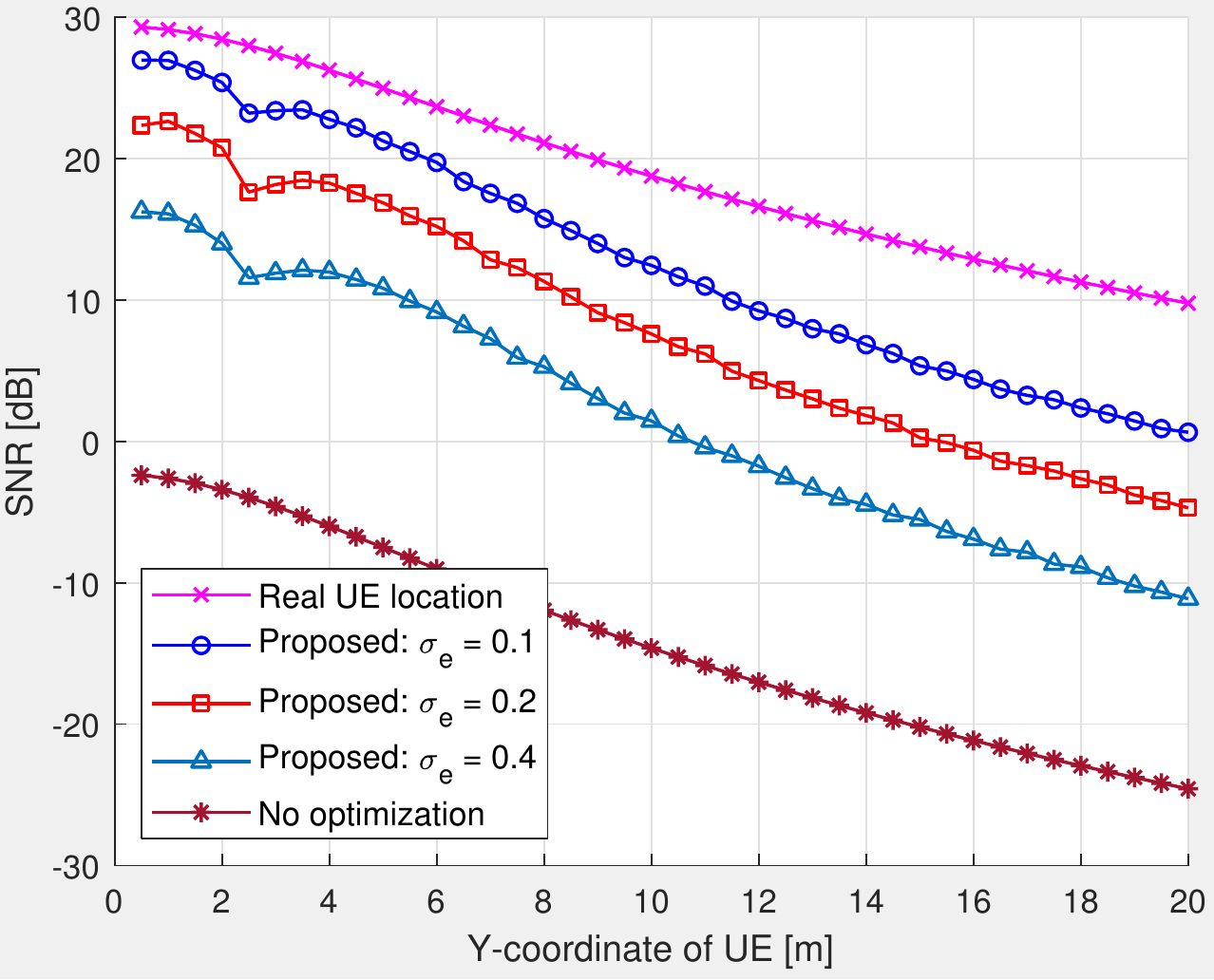}
\caption{Recived SNR vs. Y-coordinate of UE}
\label{fig:sim y snr}
\end{figure}

We present two simulation settings related to UE locations on the $x$-$y$ plane: in Fig. \ref{fig:sim x snr} and Fig. \ref{fig:sim x err}, UE's location is changed from $(0.5, 3, 0)$ to $(20, 3, 0)$ along the direction of the $x$-axis with a step of 0.5 meters.
Fig. \ref{fig:sim y snr} is the case when UE's position moves from $(5, 0.5, 0)$ to $(5, 20, 0)$ along the $y$-axis direction.

\begin{table}[!t]
\renewcommand{\arraystretch}{1.3}
\caption{The Parameters of Simulation}
\label{tab:paras}
\centering
\begin{tabular}{c|c|l}
\hline
\bfseries Symbol & \bfseries Value & \bfseries Definition \\
\hline $N$ & 8192 & Number of reflecting units on IRS\\
\hline $N_{v}$ & 64 & Number of rows of reflecting units on IRS\\
\hline $N_{h}$ & 128 & Number of volumns of reflecting units on IRS\\
\hline $D_{v}/m$ & 0.005 & Row spacing between adjacent units\\
\hline $D_{h}/m$ & 0.005 & Column spacing between adjacent units\\
\hline $M$ & 16 & Number of reflecting units on an RUS \\
\hline $N_{M}$ & 5 & Number of RUSs on the IRS\\
\hline $M_{v}$ & 4 & Number of rows of reflecting units on RUS\\
\hline $M_{h}$ & 4 & Number of columns of reflecting units on RUS\\
\hline $K$ & 128 & Number of the subbands\\
\hline $F_{c}/GHz$ & 28 & Center frequency\\
\hline $F_{d}/MHz$ & 3.6 & Bandwidth of a subband (5 RBs)\\
\hline $\alpha$ & 2 & Constant in path-loss model\\
\hline $\gamma$ & 2 & Path-loss exponent\\
\hline $\sigma^2$ & 0.001 & Power of thermal noise [Watt]\\
\hline
\end{tabular}
\end{table}


Fig. \ref{fig:sim x snr} shows the received SNR at UE side with the aid of IRS as a function of the X-coordinate of UE. The received SNR is defined as
\begin{equation}\label{eq:SNR}
\mathrm{SNR}=\frac{|(\bm{g}\odot\bm{h})\bm{\theta}|^2}{\sigma^2}.
\end{equation}
The curve ``Real UE Location" means the location of UE is perfectly known and so is the accurate channel. It is an upper bound of the performance. The curve ``No optimization" denote the case when the reflecting coefficients are fixed to 1. The three curves in the middle are the performances of our proposed CSI acquisition method with different channel estimation qualities. We may observe that when $\sigma_e = 0.1$, our proposed method has an average gain of 27 dB compared with the case of fixed reflecting coefficients. We also observe that the performance will drop when then channel estimation quality degrades. Fig. \ref{fig:sim x err} shows the error of UE location estimation, defined as $e=\|\widehat{\bm{P}}-\bm{P}\|$, under the same setting as in Fig. \ref{fig:sim x snr}. 
We can observe the location estimation error tends to increase when the UE moves far away from the IRS.

Fig. \ref{fig:sim y snr} plots the received SNR as a function of UE's Y-coordinate. We may notice a similar result as in Fig. \ref{fig:sim x snr}. The average SNR gain in the case of $\sigma_e = 0.1$ is also around 27 dB compared with the case when the phase shifts are fixed for all reflecting units.


\section{Conclusions}
This paper addressed the practical challenge of blockage in mmWave communication systems and studied the possibility of enhancing the coverage of mmWave using IRS.
We proposed an efficient CSI acquisition method for the IRS-aided communication system.  Based on the sparse nature of the mmWave channel and the large size of the IRS, we showed the possibility of learning the channel information with very few training resources using our proposed method. Simulation results demonstrated large performance gain in terms of the coverage enhancement.

\bibliographystyle{IEEEtran}
\bibliography{allCitations}

\begin{thebibliography}{10}
\providecommand{\url}[1]{#1}
\csname url@samestyle\endcsname
\providecommand{\newblock}{\relax}
\providecommand{\bibinfo}[2]{#2}
\providecommand{\BIBentrySTDinterwordspacing}{\spaceskip=0pt\relax}
\providecommand{\BIBentryALTinterwordstretchfactor}{4}
\providecommand{\BIBentryALTinterwordspacing}{\spaceskip=\fontdimen2\font plus
\BIBentryALTinterwordstretchfactor\fontdimen3\font minus
  \fontdimen4\font\relax}
\providecommand{\BIBforeignlanguage}[2]{{%
\expandafter\ifx\csname l@#1\endcsname\relax
\typeout{** WARNING: IEEEtran.bst: No hyphenation pattern has been}%
\typeout{** loaded for the language `#1'. Using the pattern for}%
\typeout{** the default language instead.}%
\else
\language=\csname l@#1\endcsname
\fi
#2}}
\providecommand{\BIBdecl}{\relax}
\BIBdecl

\bibitem{F.Boccardi:2014Five}
F.~{Boccardi}, R.~W. {Heath}, A.~{Lozano}, T.~L. {Marzetta}, and P.~{Popovski},
  ``Five disruptive technology directions for {5G},'' \emph{IEEE Commun. Mag.},
  vol.~52, no.~2, pp. 74--80, Feb. 2014.

\bibitem{TS:2013MmwaveWork}
T.~S. {Rappaport}, S.~{Sun}, R.~{Mayzus}, H.~{Zhao}, Y.~{Azar}, K.~{Wang},
  G.~N. {Wong}, J.~K. {Schulz}, M.~{Samimi}, and F.~{Gutierrez}, ``Millimeter
  wave mobile communications for {5G} cellular: It will work!'' \emph{IEEE
  Access}, vol.~1, pp. 335--349, 2013.

\bibitem{marco:2019metasuface}
M.~D. Renzo, M.~Debbah, D.~T.~P. Huy, A.~Zappone, M.~Alouini, C.~Yuen,
  V.~Sciancalepore, G.~C. Alexandropoulos, J.~Hoydis, H.~Gacanin, J.~de~Rosny,
  A.~Bounceur, G.~Lerosey, and M.~Fink, ``Smart radio environments empowered by
  {AI} reconfigurable meta-surfaces: An idea whose time has come,''
  \emph{CoRR}, vol. abs/1903.08925, 2019.

\bibitem{wuqq:2019IRS}
Q.~{Wu} and R.~{Zhang}, ``Intelligent reflecting surface enhanced wireless
  network via joint active and passive beamforming,'' \emph{IEEE Trans.
  Wireless Commun.}, pp. 1--1, Aug. 2019.

\bibitem{Tang:2019meta}
W.~{Tang}, X.~{Li}, J.~Y. {Dai}, S.~{Jin}, Y.~{Zeng}, Q.~{Cheng}, and T.~J.
  {Cui}, ``Wireless communications with programmable metasurface: Transceiver
  design and experimental results,'' \emph{China Communications}, vol.~16,
  no.~5, pp. 46--61, May. 2019.

\bibitem{Hu:2018LIS}
S.~{Hu}, F.~{Rusek}, and O.~{Edfors}, ``Beyond massive {MIMO}: The potential of
  positioning with large intelligent surfaces,'' \emph{IEEE Trans. Signal
  Process.}, vol.~66, no.~7, pp. 1761--1774, Apr. 2018.

\bibitem{Huang:2018LIS}
C.~{Huang}, G.~C. {Alexandropoulos}, A.~{Zappone}, M.~{Debbah}, and C.~{Yuen},
  ``Energy efficient multi-user {MISO} communication using low resolution large
  intelligent surfaces,'' in \emph{2018 IEEE Globecom Workshops (GC Wkshps)},
  Dec. 2018, pp. 1--6.

\bibitem{He:2019EstLIS}
\BIBentryALTinterwordspacing
Z.~He and X.~Yuan, ``Cascaded channel estimation for large intelligent
  metasurface assisted massive {MIMO},'' \emph{CoRR}, vol. abs/1905.07948,
  2019. [Online]. Available: \url{http://arxiv.org/abs/1905.07948}
\BIBentrySTDinterwordspacing

\bibitem{zheng:2019intelligent}
\BIBentryALTinterwordspacing
B.~Zheng and R.~Zhang, ``Intelligent reflecting surface-enhanced {OFDM}:
  Channel estimation and reflection optimization,'' 2019. [Online]. Available:
  \url{https://arxiv.org/abs/1909.03272}
\BIBentrySTDinterwordspacing

\bibitem{3gpp:38.211}
3GPP, \emph{{NR}; Physical Channels and Modulation (Release 15)}.\hskip 1em
  plus 0.5em minus 0.4em\relax Technical Specification TS 38.211, available:
  http://www.3gpp.org.

\bibitem{3gpp:38.901}
------, \emph{Study on Channel Model for Frequencies from 0.5 to 100 GHz
  (Release 15)}.\hskip 1em plus 0.5em minus 0.4em\relax Technical Report TR
  38.901, available: http://www.3gpp.org.

\bibitem{yin:13a}
H.~Yin, D.~Gesbert, M.~Filippou, and Y.~Liu, ``A coordinated approach to
  channel estimation in large-scale multiple-antenna systems,'' \emph{IEEE J.
  Sel. Areas Commun.}, vol.~31, no.~2, pp. 264--273, Feb. 2013.

\bibitem{Robert:2016mmwave}
R.~W. {Heath}, N.~{Gonz{\'a}lez-Prelcic}, S.~{Rangan}, W.~{Roh}, and A.~M.
  {Sayeed}, ``An overview of signal processing techniques for millimeter wave
  {MIMO} systems,'' \emph{IEEE J. Sel. Topics Signal Process.}, vol.~10, no.~3,
  pp. 436--453, Apr. 2016.

\bibitem{wang:2013triang}
Y.~{Wang}, {Xu Yang}, {Yutian Zhao}, {Yue Liu}, and L.~{Cuthbert}, ``Bluetooth
  positioning using {RSSI} and triangulation methods,'' in \emph{2013 IEEE 10th
  Consumer Communications and Networking Conference (CCNC)}, Jan 2013, pp.
  837--842.

\bibitem{3gpp:38.214}
3GPP, \emph{5G; {NR}; Physical Layer Procedures for Data (Release 15)}.\hskip
  1em plus 0.5em minus 0.4em\relax Technical Specification TS 38.214,
  available: http://www.3gpp.org.

\bibitem{EH:2012bartlett}
E.~H. {Gismalla} and E.~{Alsusa}, ``On the performance of energy detection
  using bartlett's estimate for spectrum sensing in cognitive radio systems,''
  \emph{IEEE Trans. Signal Process.}, vol.~60, no.~7, pp. 3394--3404, Jul.
  2012.

\bibitem{R.schmidt:1986music}
R.~{Schmidt}, ``Multiple emitter location and signal parameter estimation,''
  \emph{IEEE Trans. Antennas Propag.}, vol.~34, no.~3, pp. 276--280, Mar. 1986.

\bibitem{Nikolic:2013music}
\BIBentryALTinterwordspacing
M.~Nikoli\'{c}, D.~P. Jovanovi\'{c}, Y.~L. Lim, K.~Bertling, T.~Taimre, and
  A.~D. Raki\'{c}, ``Approach to frequency estimation in self-mixing
  interferometry: multiple signal classification,'' \emph{Appl. Opt.}, vol.~52,
  no.~14, pp. 3345--3350, May. 2013. [Online]. Available:
  \url{http://ao.osa.org/abstract.cfm?URI=ao-52-14-3345}
\BIBentrySTDinterwordspacing

\bibitem{roy:1989esprit}
R.~{Roy} and T.~{Kailath}, ``{ESPRIT}-estimation of signal parameters via
  rotational invariance techniques,'' \emph{{IEEE} Trans. Acoust., Speech,
  Signal Process.}, vol.~37, no.~7, pp. 984--995, Jul. 1989.

\bibitem{Duofang:2008esprit}
C.~{Duofang}, C.~{Baixiao}, and Q.~{Guodong}, ``Angle estimation using {ESPRIT}
  in {MIMO} radar,'' \emph{Electronics Letters}, vol.~44, no.~12, pp. 770--771,
  Jun. 2008.

\end{thebibliography}

\end{document}